\documentclass{article}
\usepackage{spconf,amsmath,graphicx}

\usepackage{booktabs}
\usepackage{url}
\usepackage{cite}

\usepackage{amssymb}
\usepackage{mathtools, nccmath}

\usepackage{multirow}  

\title{Deep Audio Zooming: Beamwidth-Controllable Neural Beamformer}
\name{Meng Yu, Dong Yu}
\address{Tencent AI Lab, Bellevue, WA, USA\\
      \{raymondmyu, dyu\}@global.tencent.com}
\begin{document}
%
\maketitle
\begin{abstract}
Audio zooming, a signal processing technique, enables selective focusing and enhancement of sound signals from a specified region, attenuating others. While traditional beamforming and neural beamforming techniques, centered on creating a directional array, necessitate the designation of a singular target direction, they often overlook the concept of a field of view (FOV), that defines an angular area. In this paper, we proposed a simple yet effective FOV feature, amalgamating all directional attributes within the user-defined field. In conjunction, we've introduced a counter FOV feature capturing directional aspects outside the desired field. Such advancements ensure refined sound capture, particularly emphasizing the FOV's boundaries, and guarantee the enhanced capture of all desired sound sources inside the user-defined field. The results from the experiment demonstrate the efficacy of the introduced angular FOV feature and its seamless incorporation into a low-power subband model suited for real-time applications.

\end{abstract}

\vspace{-0.2cm}
\section{Introduction}
\vspace{-0.2cm}
Audio zooming technology refers to a signal processing technique that allows a user to selectively focus on and enhance the audio signals originating from a specific region of interest in a sound field, while attenuating signals from other directions. 
It finds applications in various fields, including teleconferencing, surveillance, broadcasting and video filming. 
With the audio zooming feature, people can adjust the video recording's focus using a pinch motion on the screen. When you zoom in, the audio from your focal point becomes more pronounced. Conversely, as you zoom out, the ambient sounds become more evident and are no longer diminished.
Beamforming is the most closely related field of study to this concept,
that amplifies the sound from the target direction while suppressing all other sounds \cite{Van88, Benesty08}.
Most beamforming techniques share the same objective of improving the sound quality from a single direction. However, they do not take into account the concept of a field of view (FOV). 
The width of the enhanced beam in beamforming is determined by several factors, including the microphone configuration, signal frequency, and type of beamformer used. 
Duong et al. created the audio zoom effect by weighted mixing the beamforming enhanced target sound source and the original microphone signal \cite{Duong17}.
Nair et al. \cite{Nair19} proposed a beamforming method that depends on the sound spectral covariance matrices, taking into account the desired audio signals both within and outside the FOV. 
With sound reflections, audio waves from a source outside the FOV can reach the microphone after bouncing off surfaces from within the FOV. Under these circumstances, their audio zooming technique would continue to amplify these reflected sound signals.

An array of microphones carry spatial information about the source of a sound.
The effectiveness of established spatial features, such as inter-channel phase difference (IPD), has been demonstrated at the input stage for speech separation methods that use time-frequency (T-F) masking \cite{chen2018multi, wang2018multi, lianwu2019multi}. Additionally, to further improve the extraction of source signals from a specific direction, meticulously crafted directional features indicating the dominant directional source in each T-F bin have been introduced in prior works such as \cite{chen2018multi, wang2019combining, gu2019neural,bahmaninezhad2019comprehensive, Xu21}.
These directional features are contingent upon a specific direction of interest. Drawing inspiration from the multi-look/zone neural beamformer \cite{Yu20,Xu23},
we introduced a FOV feature that consolidates all directional features within the desired field of interest. Simultaneously, we integrate a counter FOV feature to signify all the directional features outside the angular field. Employing both these features refines sound capture, especially around the FOV's edges. 
When this feature is input into the neural network, it ensures the capture of all sound sources or speakers within the designated angular field, while attenuating those outside this zone. 
The rest of the paper is organized as follows. In Section \ref{sec:az}, we first recap the directional feature mentioned in the recent literature, and then present the new audio zooming FOV feature and its usage in the neural network model. We describe our experimental setups and evaluate the effectiveness of the proposed method in Section \ref{sec:exp}. We conclude this work in Section \ref{sec:con}.

\vspace{-0.2cm}
\section{Deep Audio Zooming Feature and Model}\label{sec:az}
\vspace{-0.2cm}
Previous work in \cite{chen2018multi,gu2019neural,bahmaninezhad2019comprehensive,Gu20} have proposed to leverage a proper designed directional feature of the target speaker to perform the target speaker separation. 
Through a short-time-Fourier-transform (STFT), the $M$ microphone signals $y_m$ is transformed to its complex spectrum $Y_m$, where $m=1, 2 \dots, M$. IPD is computed by the phase difference between channels of complex spectrograms as $\text{IPD}^{(m)}(t,f)=\angle\mathbf{Y}^{m_1}(t,f)-\angle\mathbf{Y}^{m_2}(t,f)$,
where $m_1$ and $m_2$ are two microphones of the $m$-th microphone pair out of $\overline{M}$ selected microphone pairs.
A directional feature is incorporated as a target speaker bias. This feature was originally introduced in \cite{chen2018multi}, which computes the averaged cosine distance between the target speaker steering vector and IPD on all selected microphone pairs as
\vspace{-0.15cm}
\begin{equation}\label{df}
\vspace{-0.2cm}
\text{d}_\theta(t, f)=\overset{\overline{M}}{\underset{m=1}{\sum}}
\left <\mathbf{e}^{\angle\mathbf{v}_{\theta}^{(m)}(f)},
\mathbf{e}^{\text{IPD}^{(m)}(t,f)} \right >
\end{equation}

where $\angle\mathbf{v}_{\theta}^{(m)}(f):=2\pi f\varDelta^{(m)}  \cos{\theta^{(m)}} / c$ is phase of the steering vector from the direction $\theta$ at frequency \emph{f} with respect to $m$-th microphone pair, $\varDelta^{(m)}$ is the distance between the $m$-th microphone pair, $c$ is the sound velocity, and vector $\mathbf{e}^{(\cdot)}:= [\cos(\cdot), \sin(\cdot)]^T$. If the T-F bin $(t,f)$ is dominated by the source from $\theta$, then $d_{\theta}(t,f)$ will be close to 1, otherwise it deviates towards -1. As a result, $\text{d}_\theta(t,f)$ indicates if a speaker from a desired direction $\theta$ dominates in each T-F bin. 
Such location-based input features (LBI) are found in \cite{chen2018multi, wang2019combining, gu2019neural,bahmaninezhad2019comprehensive}.

\begin{figure}[t]
\vspace{-2cm}
\hspace{-0.2cm}
  \includegraphics[width=110mm, height=90mm]{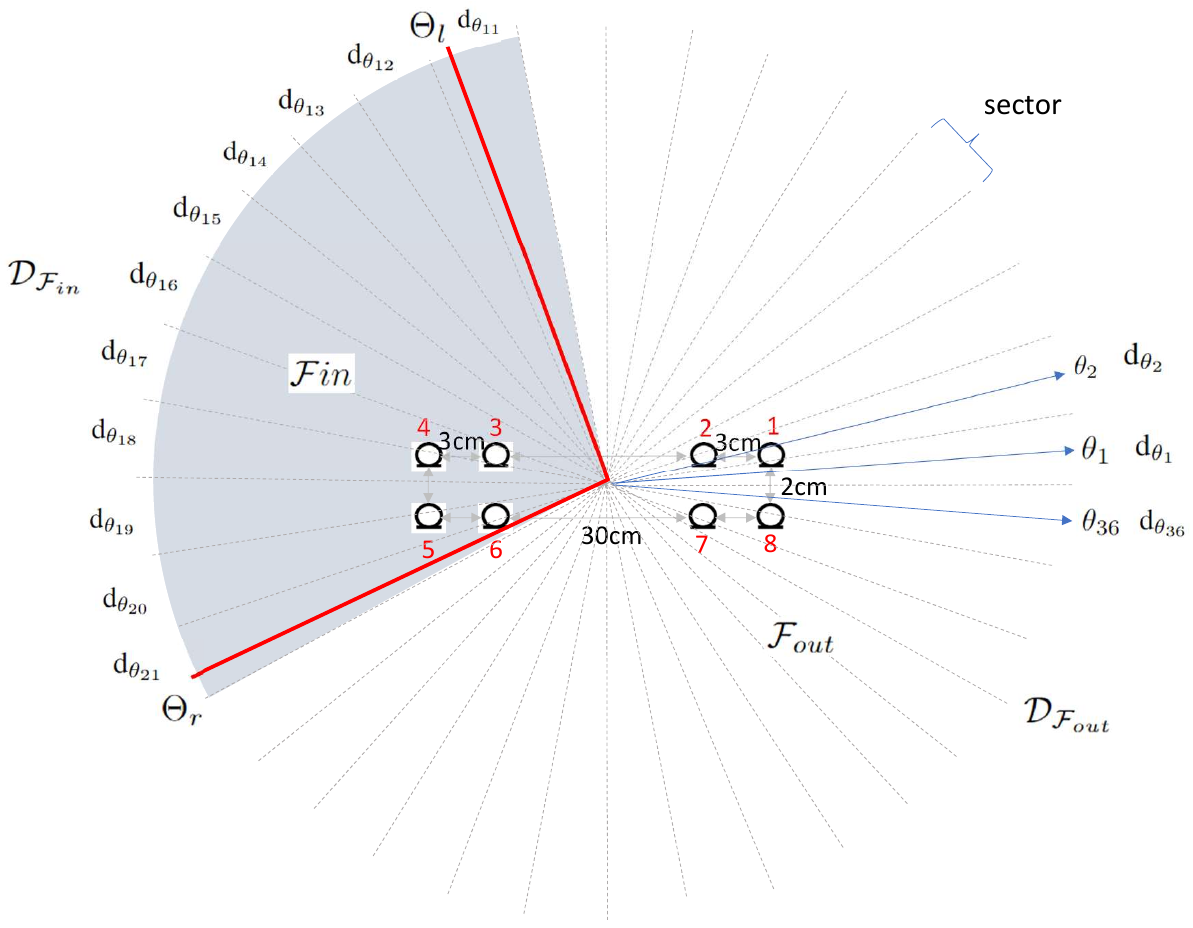}
  \vspace{-2.2cm}
 \caption{The audio zooming 2D FOV feature extraction.}
  \label{fig:fea}
 \vspace{-0.5cm}
\end{figure}

\vspace{-0.2cm}
\subsection{Audio Zooming FOV Feature}\label{subsec:mlen}
\vspace{-0.2cm}
Beyond the LBI approach, when a scenario involves multiple speakers that need to be distinctly separated, the location-based training (LBT) method is utilized \cite{Taherian22, Taherian22J}.  
For specific application scenarios, DNNs have been trained to provide distinct outputs for every designated direction, including those devoid of source activity \cite{Chazan19, Bohlender21}.
When we evenly divide the space into several sectors and sample a direction at the bisector of each sector, we establish the idea of multi-look directions. The idea of ``multi-look direction'' has been applied to speech separation \cite{Ji20,Chen17b,Chen18a, Yu20} and multi-channel acoustic model \cite{Sainath15,Sainath16, Sainath17}, respectively, where a small number of spatial look directions cover
all possible target speaker directions.

A set of $K$ directions in the horizontal plane is sampled. For example, in Fig. \ref{fig:fea}, K = 36 and the horizontal space is evenly partitioned to 36 sectors. The azimuths of look directions $\theta_{1,2,...,36} = \{5^\circ,15^\circ,\dots, 345^\circ, 355^\circ\}$, corresponding to the sectors $[0^\circ,10^\circ]$, $[10^\circ,20^\circ]$, $\dots$, $[340^\circ,350^\circ]$, and $[350^\circ,360^\circ]$, respectively. 
The look directions $\theta_{1,2,...,K}$ result in $K$ directional feature vectors $d_{\theta_k}, k=1,2,...,K$.
Obviously, $\text{d}_{\theta_k}(t, f)$ is determined by the actual source direction $\theta$ and look direction $\theta_k$. Consequently, for those T-F bins influenced predominantly by the source near the look direction $\theta_k$, the value of $\text{d}_{\theta_k}$ will be substantial.

\begin{figure}[t]
\vspace{-2cm}
\hspace{0.2cm}
  \includegraphics[width=110mm, height=85mm]{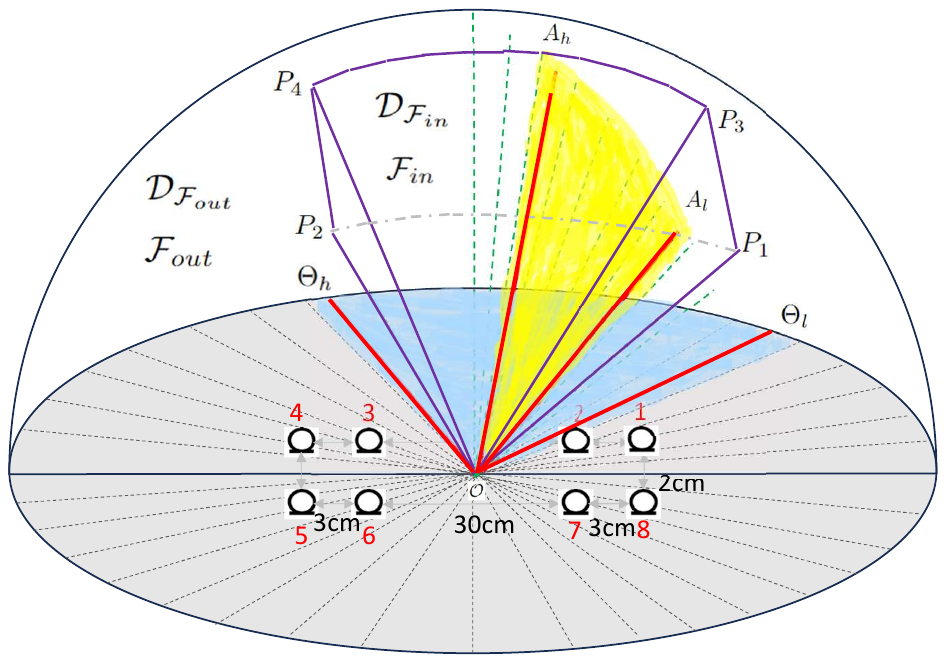}
  \vspace{-3.5cm}
 \caption{The audio zooming 3D FOV feature extraction.}
  \label{fig:fea3D}
 \vspace{-0.4cm}
\end{figure}

\begin{figure*}[!t]
        \centering
        \vspace{-5.7cm}
        \includegraphics[width=240mm,height=180mm]{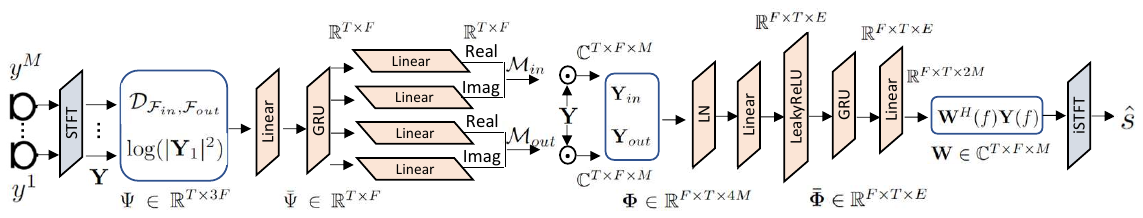}
        \vspace{-11cm}
        \caption{Deep audio zooming model diagram. Trained on 4-second chunks with the Adam optimizer and a batch size 16 for 30 epochs. The initial learning rate is set to 1e-4 with a gradient norm clipped with max norm 10.}
        \vspace{-0.3cm}
        \label{model_diagram}
\end{figure*}

For a user-specified FOV delineated by the angular boundaries $[\Theta_l, \Theta_h]$, we first determine the corresponding sector of these boundaries. The sectors on and within the range $[\Theta_l, \Theta_h]$ are represented by its index set $\mathcal{F}{in}$, i.e. the gray area in Fig.\ref{fig:fea}. On the other hand, sectors that fall outside these boundaries are represented by $\mathcal{F}_{out}$. Each of the directional feature $\text{d}_{\theta_k}$ belongs to $\mathbb{R}^{T\times F}$, where $T$ and $F$ are the total number of frames and frequency bands of the complex spectrogram, respectively. By concatenating the values of $\text{d}_{\theta_k}$ where $ k \in \mathcal{F}_{in}$, the resulting dimension becomes considerably large. Moreover, this dimension fluctuates with variations in the field size, making it unsuitable for model training. 
We introduce a method to consolidate the high-dimensional directional feature sets by employing a max operation across the sectors, as detailed in 
\vspace{-0.2cm}
\begin{equation}\label{fea_1}
    \mathcal{D}_{\mathcal{F}_{in}}(t, f) = \mathop{\max}_{\theta_k \in \mathcal{F}_{in}} \text{d}_{\theta_k}(t, f).
    \vspace{-0.2cm}
\end{equation}
The underlying idea is that when a speaker resides within the field $\mathcal{F}_{in}$, there will inevitably be a sampled direction, $\theta_{k_0}, k_0 \in \mathcal{F}_{in}$, nearer to this speaker than other sampled directions. This results in $\text{d}_{\theta_{k_0}}$ being the most prominent at the T-F bins where this speaker has dominance. Consequently, employing the max operation for $\text{d}_{\theta_{k}}$ within $\mathcal{F}_{in}$ can effectively capture the prominence of all active speakers within the targeted field. Meanwhile, following the max operation, the dimension of the resulting feature vector $\mathcal{D}_{\mathcal{F}_{in}}$ remains consistent with that of any original directional feature $\text{d}_{\theta_k}$.
At the same time, we incorporate a counter FOV feature to represent all directional attributes beyond the angular field. Using both these features sharpens the sound capture, particularly near the boundaries of the FOV. This enhances the model's ability to differentiate between sound sources within and outside the targeted field. The counter FOV feature $ \mathcal{D}_{\mathcal{F}_{out}}(t, f)$ is computed similarly as Eq. \ref{fea_1}, with $\theta_k \in \mathcal{F}_{out}$. 
There are two methods to further consolidate the two FOV feature vectors. The first approach involves concatenation, represented as 
\begin{equation}\label{cat_fea}
    \mathcal{D}_{\mathcal{F}_{in},\mathcal{F}_{out}}= [\mathcal{D}_{\mathcal{F}_{in}}, \mathcal{D}_{\mathcal{F}_{out}}] \in \mathbb{R}^{T\times 2F }.
\end{equation}
The second is through post-processing, as calculated in Eq. \ref{post},
\begin{equation}\small\label{post}
    \mathcal{D}_{\mathcal{F}_{in},\mathcal{F}_{out}}(t, f) = \begin{cases}
         -1 &\text{if $\mathcal{D}_{\mathcal{F}_{in}}(t, f) \leq \mathcal{D}_{\mathcal{F}_{out}}(t, f)$} \\
         \mathcal{D}_{\mathcal{F}_{in}}(t, f) &\text{else}
    \end{cases}
\end{equation}
The post-processing aims to remove the components of speakers outside the desired region, thereby enhancing the prominence of the speakers within the region. 

To adapt the aforementioned feature extraction to a 3D space \cite{Gu21}, we merely modify the directional feature $\text{d}_\theta$ in Eq. \ref{df} such that $\angle\mathbf{v}_{\theta}^{(m)}(f):=2\pi f\varDelta^{(m)}  \cos{\theta^{(m)}} \cos{\alpha^{(m)}}/ c$, with $\alpha$ representing the elevation angle in the vertical dimension. Concurrently, the definitions of $\mathcal{F}_{in}$ and $\mathcal{F}_{out}$ will also account for the preferred elevation span. As illustrated in Fig. \ref{fig:fea3D}, two red lines at angles $\Theta_l$ and $\Theta_h$ delineate the desired angular boundary on the horizontal plane. Simultaneously, two red lines at angles $A_u$ and $A_d$ set the boundary on the vertical plane. 
When considering these angular boundaries in a 3D space collectively, the sectors within the field, denoted as $\mathcal{F}_{in}$, shape the pyramid defined by $\overline{\mathcal{O}P_1P_2P_3P_4}$.

\vspace{-0.2cm}
\subsection{Subband Model}\label{sec:model}
\vspace{-0.2cm}
We formulate the problem as amplifying speech signals from a specified angular range $[\Theta_l, \Theta_h]$ and $[A_l, A_h]$. 
The diagram of the deep audio zooming model, which is similar yet more optimized than the model presented in \cite{Xu21, Xu23}, can be seen in Fig. \ref{model_diagram}. By computing STFT (512 window size and 256 hop size) on $M=8$ microphone channels, a reference channel, e.g. the first channel complex spectrogram $\mathbf{Y}_1$, is used to compute logarithm power spectrum (LPS) by $\log(|\mathbf{Y}_1|^2) \in \mathbb{R}^{T\times F}$. The spectral feature LPS, combined with the introduced audio zooming FOV feature $\mathcal{D}_{\mathcal{F}_{in},\mathcal{F}_{out}}$ in Eq. \ref{cat_fea}, are utilized as the input for the neural network. 
Subsequent to processing through a GRU layer, the model deduces $T\times F$ dimensional complex-valued masks: $\mathbf{\mathcal{M}}_{in}(t,f)$ and $\mathbf{\mathcal{M}}_{out}(t,f)$ for source estimations within and outside the FOV, respectively, via four linear layers. Noise signals are associated with the sources outside the FOV since noise reduction is integral to our objective. The two masks are multiplied with each microphone spectrum $\mathbf{Y}^m$ to estimate the target signals and interfering signals at each microphone $m$.
$M$ channels of real and imaginary parts of $\mathbf{Y}_{in}$ and $\mathbf{Y}_{out}$ are then concatenated, represented as $\mathbf{\Phi}\in \mathbb{R}^{F\times T \times 4M}$. 
After layer normalization \cite{Ba16} and a linear projection with leakyRELU activation, a RNN layer together with a linear layer receive the subband embedding of individual frequency bands $\mathbf{\bar{\Phi}}(f)\in \mathbb{R}^{T \times E}$, where $f=1,2,\dots F$ and $E=32$ is the dimension of the subband embedding, to estimate the enhancement filters at the corresponding frequency as $\mathbf{W}(f)\in \mathbb{C}^{T \times M}$. We compute the output signal $\hat{\mathbf{S}}$ by $\mathbf{W}^H(f) \mathbf{Y}(f)$. 
The total number of parameters is 860K and the computation cost is 184 MMACs/s. Same as the one in \cite{Yu22}, the loss function is calculated using the combination of negative SISDR\cite{Roux19} in time domain and $l_1$ norm on the spectrum magnitude difference. 

\vspace{-0.2cm}
\section{Experiments and Results} \label{sec:exp}
\vspace{-0.2cm}
We simulate the 8-channel dataset using
AISHELL-2 \cite{Du2018}. The geometry of the microphone array, which is mounted on the ceiling, is depicted in Fig. \ref{fig:fea}. We generate 10K room impulse responses with random room characteristics 
and reverberation time ranging from 0.3s to 1.3s using gpuRIR \cite{Diaz20}. In addition, environmental noises are added with signal-to-noise-ratio ranging from 10 to 40dB. 95K, 2.5K, and 100 utterances are generated for training, validation and testing, respectively. A total of up to five speakers are randomly sampled.
For each training utterances we randomly sample the angular range $[\Theta_l, \Theta_h]$ and $[A_l, A_h]$, ensuring the model encounters a variety of speaker counts both within and outside the designated target field.

\begin{figure}[t]
\vspace{-1.9cm}
\hspace{0.3cm}
  \includegraphics[width=180mm, height=80mm]{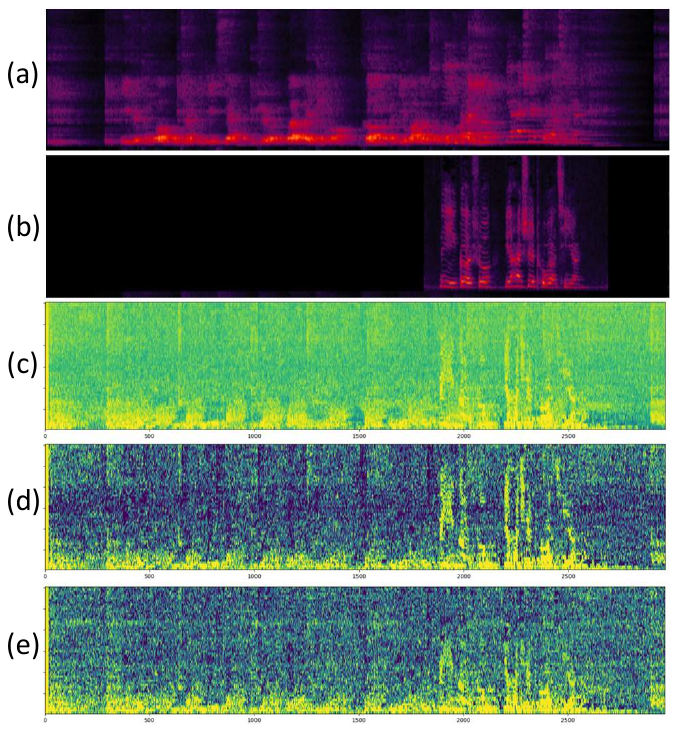}
  \vspace{-2.9cm}
 \caption{Audio zooming FOV feature extraction. a) microphone signal, b) speaker in the target region, c) original feature $\mathcal{D}_{\mathcal{F}_{in}}$ in Eq. \ref{fea_1}, d) post processed feature $\mathcal{D}_{\mathcal{F}_{in},\mathcal{F}_{out}}$ in Eq. \ref{post}. e) feature in Eq. \ref{post} with low resolution.  }
  \label{fig:feapost}
\end{figure}

We begin by examining the efficacy of the introduced audio zooming FOV feature, $\mathcal{D}_{\mathcal{F}{in},\mathcal{F}_{out}}$, in its two representations Eq. \ref{cat_fea} and \ref{post}. The horizontal and vertical resolutions, which correspond to the sector width, are configured at $20^\circ$ and $10^\circ$, respectively. The microphone pairs chosen to compute IPD are (1, 4), (2, 6), (1, 7), (2, 7), (4, 6), and (3, 7). 
For every test sample, the beam-center aligns with the angular direction of one speaker. Meanwhile, the beam-width undergoes random sampling on both the horizontal and vertical planes. Consequently, the target output signal always contains at least one speaker. Fig. \ref{fig:feapost} displays the extracted feature for a target area with an active speaker. Comparing the raw audio zooming feature from Eq. \ref{fea_1} with the post-processed feature in Eq. \ref{post}, it's evident that the post-processing refines the feature pattern. This refinement results in the attenuation of T-F bins dominated by sound sources outside the target region, thereby accentuating the speakers within the desired area and enhancing their distinctiveness. The findings in Table \ref{eval_1} indicate that concatenating the FOV feature from both inside and outside the target region (``our best'') is more effective than using the post-processed version. This discrepancy arises because the "winner-take-all" approach in post-processing doesn't cater well to bins where speakers overlap in the T-F domain. As a result, the post-processed feature might lead to attenuation of the target speaker in certain T-F bins.

\begin{figure}[t]
        \vspace{-2.0cm}
        \hspace{0.1cm}
        \includegraphics[width=180mm,height=140mm]{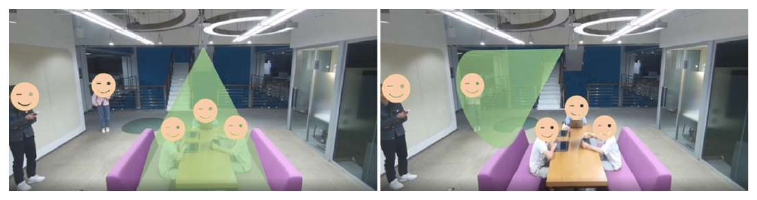}
        \vspace{-10.6cm}
        \caption{Deep audio zooming real demo. }
        \vspace{-0.5cm}
        \label{real}
\end{figure}

The calculation of the FOV feature relies on space partitioning, essentially referring to the sampling resolution of the look directions. A higher resolution leads to a more precise raw directional feature (as per Eq. \ref{df}).
Consequently, post-aggregation through the max operation detailed in Eq. \ref{fea_1}, the resultant audio zooming feature is sharpened, allowing for superior differentiation between sources within and outside the target region. 
The row (e) in Fig. \ref{fig:feapost} displays the feature extracted at the horizontal resolution $60^\circ$ and vertical resolution $15^\circ$. A more refined resolution 
demonstrates markedly enhanced feature prominence in row (d).

In our subband model, the pivotal component is the second GRU layer, which processes data for each frequency band separately. The model produces $M$ channel complex weights, and the final output is derived from the weighted summation of all microphone signals, functioning similar to beamformer filtering. For a comparative analysis, we crafted a full-band GRU model. This was achieved by eliminating the layers and operations post the layer normalization as seen in Fig. \ref{model_diagram}, augmenting the initial GRU with additional layers, and expanding its size. This ensures that, in terms of computational expense, the full-band model aligns with the subband model. The ultimate output is procured via $\mathcal{M}_{in}\textbf{Y}_1$. 
Results from this model comparison are detailed in Table \ref{eval_1}. The subband model (``our best'') distinctly surpasses the fullband model.

\begin{table}[]\small
\centering
\vspace{-0.9cm}
\caption{\label{eval_1} {\it Audio zooming experimental comparisons}}
\begin{tabular}{c|c|c|c}\toprule
 \midrule Category & Method  & PESQ \textsuperscript{\cite{ITU-T01}} &  SDR[dB]\textsuperscript{\cite{Vincent06}} \\
  \midrule - & no processing  & 2.01 & 4.94 \\
  \midrule \multirow{2}{*}{feature}  & $\mathcal{D}_{\mathcal{F}{in}}$ Eq. \ref{fea_1} & 2.66 & 10.19\\
    &$\mathcal{D}_{\mathcal{F}{in}, \mathcal{F}{out}}$ Eq. \ref{post} & 2.73 & 11.32 \\
   \midrule resolution & h.:$60^\circ$, v.: $15^\circ$  & 2.56 & 10.55\\
   \midrule model & fullband model  & 2.44 & 10.32\\
   \midrule \# of channels & 3-mic & 2.61 & 9.94\\
    \midrule \multirow{2}{*}{baseline}& oracle MVDR  &  2.31 &  6.09\\
  &$\text{d}_\theta$ Eq. \ref{df} & 2.43 & 9.10 \\
   \midrule   \multirow{4}{*}{our best}  & $\mathcal{D}_{\mathcal{F}{in}, \mathcal{F}{out}}$ Eq. \ref{cat_fea}& \multirow{4}{*}{\textbf{2.78}}  & \multirow{4}{*}{\textbf{11.50}} \\
   & h.:$20^\circ$, v.: $10^\circ$ &  & \\
   & subband &  & \\
   & 8-mic &  & \\
   \hline
   \bottomrule
   
\end{tabular}
\vspace{-0.5cm}
\end{table}

Our subsequent investigation delves into the performance implications of utilizing a significantly smaller number of microphones, such as three microphones, akin to those found on a phone. In this 3-channel array setup, microphones 1, 2, and 7 are utilized. The pairs (1,2), (1, 7), and (2, 7) are employed to compute the IPD. From Table \ref{eval_1}, it reveals that the 3-microphone model's performance is almost on par with the 8-microphone model, albeit with slight discrepancies.

The oracle MVDR, derived from the Ideal Ratio Mask (IRM), serves as our baseline. For each test utterance, we compute the IRM using the target and microphone signals. This IRM then helps compute the noise covariance matrix. We use the ground-truth beam-center as the target MVDR direction. This approach positions the oracle MVDR as the upper limit for the audio zooming method in \cite{Nair19}. Additionally, the directional feature from Eq. \ref{df} is inputted into our subband model for a direction-specific LBI enhancement, rather than a field-wide enhancement. The details of the comparison can be found in Table \ref{eval_1}.
The live demonstration under an actual scenario was carried out in an office setting, as depicted in Fig. \ref{real}. In this scene, five individuals are engaged in conversation; three of them are seated at a table, while the other two stand beside a wall at some distance from the table. The device's processing capabilities allow for user-directed control. For instance, it can focus solely on the individuals seated at the table or target a specific area away from the table. When no speaker is present in the target region, there's an attenuation of about 49.0dB. Moreover, the DNSMOS score \cite{reddy2022dnsmos} for test samples rose from 1.98 to 2.49.

\vspace{-0.2cm}
\section{Conclusion} \label{sec:con}
\vspace{-0.2cm}
In this study, we introduced a straightforward yet potent angular region feature that allows the neural network model to perform audio zooming efficiently. The methods of random field sampling during training and converting field boundaries into look directions are both essential components for achieving beamwidth-adjustable neural beamforming. The experimental results highlight the promise of our pioneering deep learning-based audio zooming approach. In subsequent studies, we plan to delve into how to more effectively utilize different microphone pairs during the feature computation.

\newpage
\bibliographystyle{IEEEbib.bst}
\bibliography{refs}

\end{document}